\documentclass[twocolumn]{revtex4-1}
\usepackage{graphicx,hyperref}
\usepackage{verbatim}
\usepackage{amsbsy}
\usepackage{amsmath}
\usepackage{color}
\usepackage{bm}
\usepackage{marvosym}
\usepackage[normalem]{ulem}
\usepackage{mhchem}
\usepackage{hyperref}
\hypersetup{
  colorlinks   = true, 
  urlcolor     = blue,
  linkcolor    = red, 
  citecolor   = green 
}

\newcommand{\ud}{\mathrm{d}}

\begin{document}

\title{Identifying materials with charge-spin physics using charge-spin susceptibility computed from first principles}

\author{J. N. B. Rodrigues$^{1,2}$ and Lucas K. Wagner$^{1}$}

\affiliation{$^{1}$ Department of Physics and Institute for Condensed Matter Theory, University of Illinois at Urbana-Champaign, Urbana, IL 61801, USA.\\$^{2}$ Centro de Ci\^encias Naturais e Humanas, Universidade Federal do ABC - UFABC, 09210-580, Santo Andr\'e, SP, Brasil.}


\date{\today}

\begin{abstract}
The authors present a quantity termed charge-spin susceptibility, which measures the charge response to spin degrees of freedom in strongly correlated materials.
This quantity is simple to evaluate using both standard density functional theory and many-body electronic structure techniques, enabling comparison between different levels of theory.
A benchmark on 28 layered magnetic materials shows that large values of charge-spin susceptibility correlate with unconventional ground states such as disordered magnets and unconventional superconductivity.
\end{abstract}

\maketitle

\section{Introduction}

The conventional paradigm of condensed matter physics involves the partitioning of electronic ground states into descriptions of spin and electronic degrees of freedom, potentially with small coupling terms between them.
Many materials fall under this paradigm, such as antiferromagnetic insulators, ferromagnetic metals, and non-magnetic insulators. 
However, this weak coupling paradigm is insufficient to describe some materials, which often show unconventional ground states and excitations.
Examples of unconventional behavior include the unconventional superconductors \cite{Scalapino_RMP:2012,Proust_ARCMP:2019}, unconventional magnetic states \cite{Fert_NatRevMat:2017,Anand_PRB:2014,Yamauchi_PRB:2015}, spin liquids \cite{Takagi_NatRevPhys:2019} and strong magnetodielectric effects \cite{Lawes_PSSC:2009}.
It is still an open question how to predict \textit{a priori} whether a material breaks the weakly coupled spin/electron paradigm, even if such coupling has been previously studied, mostly in the context of model Hamiltonians \cite{Guinea_PRB:2000,Jackeli_PhysB:1999,Pandey_PRB:2008,Johnson_PRB:2003,Zheng_PhysC:2001,Scalapino_RMP:2012}.

One of the most prominent examples of charge-spin interactions are the high-temperature unconventional superconductors like copper oxides and iron-based pnictides and chalcogenides.
As pointed out by Scalapino\cite{Scalapino_RMP:2012}, there is vast experimental evidence such as the proximity between magnetic ordered phases and superconductivity suggesting that in these materials the coupling between orbital (charge) and magnetic (spins) degrees of freedom plays a crucial role in determining the physics.

Despite recent progress in the study of model Hamiltonians commonly associated with materials showing unconventional properties \cite{Lederer_PNAS:2017,Jiang_PRB:2018,Esterlis_PRB:2018,Seki_PRB:2019,Berg_ARCMP:2019,Costa_CommPhys:2020}, the study of real materials with such properties remains challenging.
There have been several attempts to computationally predict whether specific groups of materials show unconventional electronic phases, for example unconventional superconductivity \cite{Klintenberg_CMS:2013,Yee_PRB:2014,Botana_PRM:2017}.
However, to our knowledge most of these similarity-based searches have attained limited success.
In this manuscript we adopt a somewhat different approach: we concentrate on exploring a new computational probe of charge-spin coupling in materials.
We present and test the charge-spin susceptibility which is computed from first principles in a simple way.
This quantity measures the response of a material's charge density to changes in its spin density.

We use density functional theory (DFT) and quantum Monte Carlo (QMC) calculations to estimate the charge-spin coupling in a set of 28 layered strongly correlated transition metal compounds including unconventional behavior such as unconventional superconductivity in the cuprates and iron-based superconductors \cite{Proust_ARCMP:2019,Si_NatRevMat:2016}, disordered magnetic ground states and bad metallic behavior in materials like BaCo$_{2}$As$_{2}$ and Sr$_{2}$VO$_{4}$ \cite{Anand_PRB:2014,Yamauchi_PRB:2015,Karmakar_PRL:2015}.
In order to assess the quality of the DFT results, for a small set of materials we compare the DFT-derived charge-spin susceptibility predictions with those obtained from fixed-node diffusion Monte Carlo.
We find that while DFT+U predicts a different charge-spin response from the QMC results, it is sufficient to distinguish large responses from small responses.
Using DFT+U calculations on the entire set of materials, we find that materials with a large charge-spin susceptibility often present unconventional phases, while materials in the same class with smaller charge-spin susceptibility do not.

\section{Methods}

\label{sec:methods}

\subsection{Test set}

\label{sec:testset}

In this work we concentrate on investigating a set of 28 magnetic layered materials containing transition metal atoms with magnetic moments arranged in diverse two dimensional structural motifs.
Some of these materials are well known to present unconventional  phases of matter as described in the interoduction.
In Table~\ref{tab:testset} we list the materials in our test set along with the low-temperature magnetic and electronic phases they show upon chemical doping or pressure.

\begin{table}
  \begin{center}
\begin{tabular}{||c c c c||} 
  \hline
  Material & Magnetism & Charge transport & $\chi_{cs}^{U=5}$\\ [0.2ex] 
  \hline\hline
  BaCo$_{2}$As$_{2}$ \cite{Pfisterer_ZFNB:1980} & {\color{blue}{D}} \cite{Anand_PRB:2014} & M \cite{Anand_PRB:2014} & 0.45 \\ [0.2ex]
  Sr$_{2}$VO$_{4}$ \cite{Cyrot_JSSC:1990} & {\color{blue}{D}}\cite{Yamauchi_PRB:2015} & I $\rightarrow$ M, {\color{blue}{B}} \cite{Karmakar_PRL:2015} & 0.42 \\ [0.2ex]
  T'-La$_{2}$CuO$_{4}$ \cite{Hord_PRB:2010} & N $\rightarrow$ S, {\color{blue}{D}} \cite{Jin_Nature:2011} & I $\rightarrow$ {\color{blue}{B}}, {\color{blue}{uSC}} \cite{Jin_Nature:2011} & 0.41 \\ [0.2ex]
  Sr$_{2}$CoO$_{4}$ \cite{Wang_PRB:2005} & F \cite{Wang_PRB:2005} & M $\rightarrow$ I, SC \cite{Wang_PRB:2005,Wang_PRB:2005,Shimada_PRB:2006,Yao_JAP:2012,takada_superconductivity_2003} & 0.39 \\  [0.2ex]
  o-BaFe$_{2}$As$_{2}$ \cite{Mittal_PRB:2011} & S $\rightarrow$ {\color{blue}{D}} \cite{Luo_PR:2012,Wang_PRL:2009} & M $\rightarrow$ {\color{blue}{B}}, {\color{blue}{uSC}} \cite{Wang_PRL:2009,Kasahara_PRB:2010,Rotter_PRL:2008} & 0.37 \\  [0.2ex]
  t-BaFe$_{2}$As$_{2}$ \cite{Jorgensen_EPJB:2010} & S $\rightarrow$ {\color{blue}{D}} \cite{Luo_PR:2012,Wang_PRL:2009} & M $\rightarrow$ {\color{blue}{B}}, {\color{blue}{uSC}} \cite{Wang_PRL:2009,Kasahara_PRB:2010,Rotter_PRL:2008} & 0.35 \\  [0.2ex]
  FeTe \cite{Awana_PhysC:2011} & S $\rightarrow$ {\color{blue}{D}} \cite{Martinelli_PRB:2010} & M $\rightarrow$ {\color{blue}{uSC}}, {\color{blue}{B}} \cite{Martinelli_PRB:2010,Sales_PRB:2009} & 0.32 \\  [0.2ex]
  FeS \cite{Lai_JACS:2015} & {\color{blue}{D}} $\rightarrow$ S \cite{Holenstein_PRB:2016} & M $\rightarrow$ {\color{blue}{uSC}} \cite{Lai_JACS:2015} & 0.30 \\  [0.2ex]
  t-FeSe \cite{Millican_SSC:2009} & {\color{blue}{D}} $\rightarrow$ S \cite{Wang_NatComm:2016} & {\color{blue}{uSC}} $\rightarrow$ {\color{blue}{B}}, M \cite{Sales_PRB:2009,Medvedev_NatMat:2009} & 0.26 \\  [0.2ex] 
  Sr$_{2}$FeO$_{4}$ \cite{Dann_JMC:1993} & N \cite{Dann_JMC:1993} & I $\rightarrow$ M \cite{Rozenber_PRB:1998} & 0.22 \\  [0.2ex]
  o-FeSe \cite{Millican_SSC:2009} & {\color{blue}{D}} $\rightarrow$ S \cite{Wang_NatComm:2016} & {\color{blue}{uSC}} $\rightarrow$ {\color{blue}{B}}, M \cite{Sales_PRB:2009,Medvedev_NatMat:2009} & 0.20 \\  [0.2ex]
  T-La$_{2}$CuO$_{4}$ \cite{Rial_PhysC:1997} & N $\rightarrow$ S, {\color{blue}{D}} \cite{Proust_ARCMP:2019} & I $\rightarrow$ {\color{blue}{B}}, {\color{blue}{PG}}, {\color{blue}{uSC}} \cite{Proust_ARCMP:2019} & 0.15 \\  [0.2ex]
  CaCuO$_{2}$ \cite{Schwer_SHTS:1997} & N $\rightarrow$ {\color{blue}{D}} \cite{Lombardi_PRB:1996} & I $\rightarrow$ {\color{blue}{uSC}} \cite{Lombardi_PRB:1996} & 0.14 \\  [0.2ex]
  SrCuO$_{2}$ \cite{Er_PhysC:1997} & N $\rightarrow$ {\color{blue}{D}} \cite{Zaliznyak_PRL:2004} & I $\rightarrow$ {\color{blue}{uSC}} \cite{Smith_Nat:1991} & 0.13 \\  [0.2ex]
  K$_{2}$CoF$_{4}$ \cite{Babel_ZAAC:1982} & N \cite{Breed_Physica:1969} & I \cite{Samoggia_SSC:1985} & 0.10 \\ [0.2ex]
  TaS$_{2}$ \cite{Spijkerman_PRB:1997} & {\color{blue}{D}} \cite{Kratochvilova_NPJQM:2017} & I $\rightarrow$ M, SC \cite{Sipos_NatMat:2008} & 0.09 \\ [0.2ex]
  Sr$_{2}$CrO$_{4}$ \cite{Baikie_JSSC:2007} & N \cite{Sakurai_JPSJ:2014} & I \cite{Baikie_JSSC:2007} & 0.06 \\ [0.2ex]
  BaCr$_{2}$As$_{2}$ \cite{Pfisterer_ZFNB:1980} & N \cite{Singh_PRB:2009} & M \cite{Singh_PRB:2009} & 0.05 \\ [0.2ex]
  La$_{2}$NiO$_{4}$ \cite{RodriguezCarvajal_JPCM:1991} & N $\rightarrow$ S \cite{Wochner_PRB:1998} & I \cite{RodriguezCarvajal_JPCM:1991} & 0.04 \\ [0.2ex]
  BaMn$_{2}$As$_{2}$ \cite{Brechtel_ZFNB:1978} & N \cite{An_PRB:2009} & I $\rightarrow$ M \cite{Pandey_PRL:2012} & 0.03 \\ [0.2ex]
  NiPSe$_{3}$ \cite{Brec_AC:1980} & N \cite{LeFlem_JPCS:1982} & I \cite{LeFlem_JPCS:1982} & 0.03 \\ [0.2ex]
  La$_{2}$CoO$_{4}$ \cite{Skinner_JSSC:2007} & N $\rightarrow$ S \cite{Babkevich_NatComm:2016} & I \cite{Babkevich_PRB:2010} & 0.02 \\ [0.2ex]
  Sr$_{2}$MnO$_{4}$ \cite{Tezuka_JSSC:1999} & N \cite{Kao_JMCC:2015} & I \cite{Kao_JMCC:2015} & 0.02 \\ [0.2ex]
  TeCuO$_{3}$ \cite{Philippot_RCM:1976} & N \cite{Lawes_PSSC:2009} & I \cite{Lawes_PSSC:2009} & 0.02 \\ [0.2ex]
  CrGeTe$_{3}$ \cite{Carteaux_JPCM:1995} & F \cite{Carteaux_JPCM:1995} & I \cite{Carteaux_JPCM:1995} & 0.01 \\ [0.2ex]
  K$_{2}$CuF$_{4}$ \cite{Herdtweck_ZAAC:1981} & F \cite{Yamada_JPSJ:1972} & I \cite{Kleemann_JPCSSP:1981} & 0.01 \\ [0.2ex]
  K$_{2}$NiF$_{4}$ \cite{Yeh_ACB:1993} & N  \cite{Birgeneau_PRB:1970} & I \cite{Birgeneau_PRB:1970} & 0.01 \\ [0.2ex]
  SeCuO$_{3}$ \cite{Escamilla_JSSC:2002} & F \cite{Lawes_PSSC:2009} & I \cite{Lawes_PSSC:2009} & 0.01 \\ [0.2ex]
  \hline
\end{tabular}
\begin{tabular}{c c l c||c c c l}
  & & &&& M &=& metal \\ [0.1ex]
  N &=& N\'eel order &&& B &=& bad metal \\ [0.1ex]
  S &=& Stripe order &&& PG &=& pseudogap \\ [0.1ex]
  F &=& Ferromagnet &&& SC &=& conv. supercond. \\ [0.1ex]
  D &=& Disordered &&& uSC &=& unconv. supercond. \\ [0.1ex]
  && &&& I &=& insulator \\ [0.1ex]
\end{tabular}
\end{center}
\caption{Materials in our test set ordered according to their average charge-spin susceptibility, $\chi_{cs}$, obtained from DFT+U with $U=5$~eV.
  Each material's magnetic and electronic phases are shown in the two central columns.
  The arrows stand for phases achieved by chemical doping or pressure.
}
\label{tab:testset}
\end{table}

\subsection{Charge-spin susceptibility as an estimate of charge-spin coupling}

\label{sec:chargespin}

\label{sec:calc_chgspin}
\begin{figure}
  \centering
  \includegraphics[width=0.99\columnwidth]{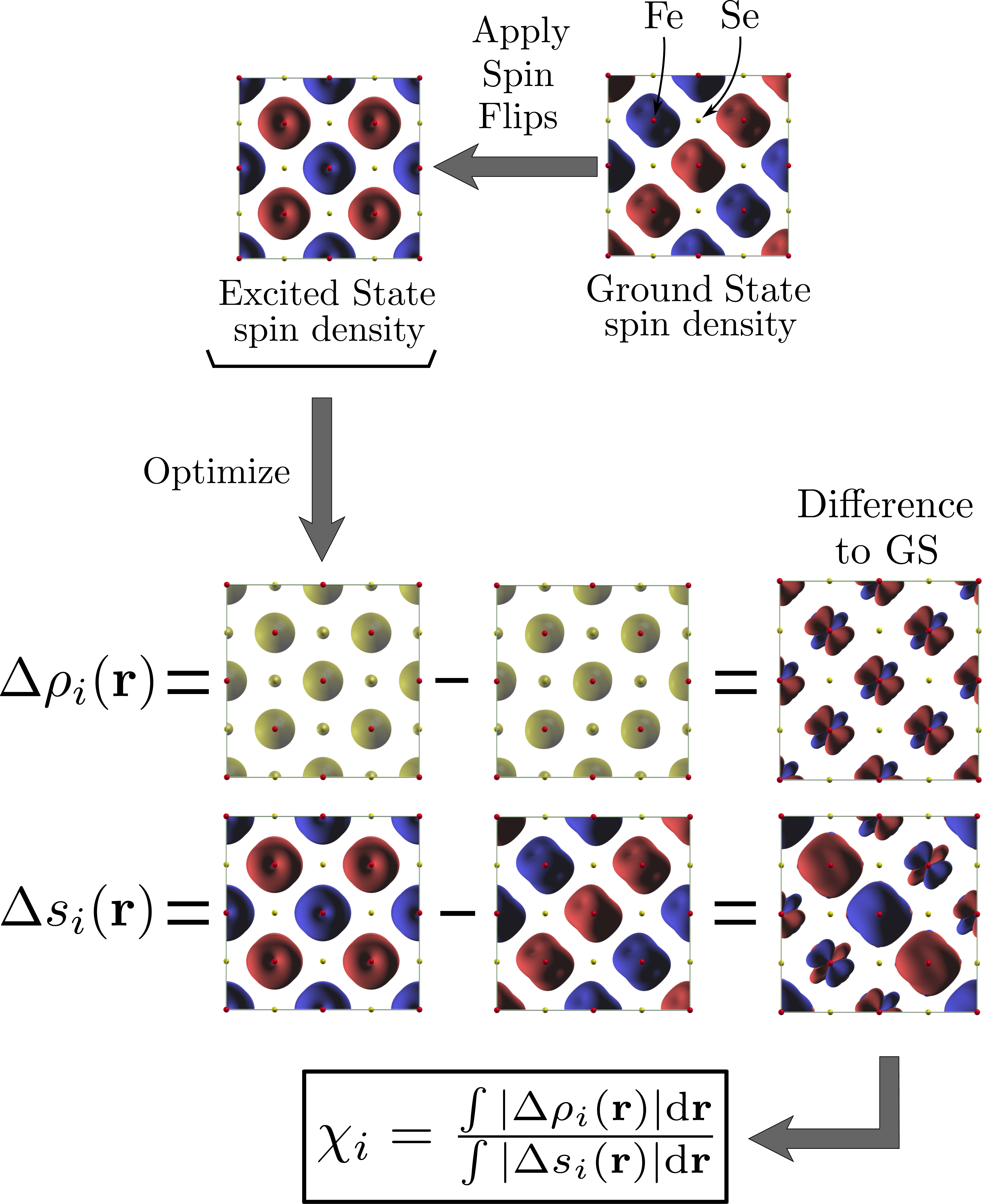}
  \caption{\textbf{Schematic representation of the methodology used to compute the charge-spin susceptibility.}
    Starting from the ground state magnetic texture of, eg. FeSe (iron atoms with stripe order), we flip some of the magnetic moments to obtain a different magnetic order (eg., N\'eel magnetic order).
    We optimize that initial texture to obtain a low-energy state with that magnetic order.
    With these two states we compute the difference between charge densities, $\Delta \rho_{i}(\mathbf{r})$, and spin densities, $\Delta s_{i}(\mathbf{r})$, from which we obtain the charge-spin susceptibility of this pair of states, $\chi_{i} = \int \vert \Delta \rho_{i}(\mathbf{r}) \vert \ud \mathbf{r} / \int \vert \Delta s_{i}(\mathbf{r}) \vert \ud \mathbf{r}$.
    In the panels above, yellow, red and blue identify isodensity surfaces in the ab-plane (of the iron atoms) for the charge density, positive and negative spin density, respectively.
    The iso-charge density surfaces (yellow) correspond to a isolevel of $0.15~e/\mathring{A}^{3}$, while all the other isodensity surfaces (red/blue) correspond to a isolevel of $0.01~e/\mathring{A}^{3}$.
  }
  \label{fig:scheme}
\end{figure}

We shall start by clarifying the relation between the charge-spin susceptibility and the coupling between charge and spin degrees of freedom in model effective Hamiltonians.
While our methodology does not depend on any particular effective Hamiltonian being applicable, we will illustrate this relation using a toy model.
Consider the effective Hamiltonian
\begin{eqnarray}
   H &=& H_{o} + H_{S} + \lambda \, H_{oS} \,, \label{eq:H}
\end{eqnarray}
where $H_{o}$ describes the orbital degrees of freedom, while $H_{S}$ describes spin states.
The term $H_{oS}$ accounts for interactions between the latter two sets of degrees of freedom, which are controlled by the coupling $\lambda$.

In order to compute the relation between the charge-spin coupling $\lambda$ and the charge-spin susceptibility in a system governed by Eq.~\ref{eq:H}, consider a small deformation of the electronic wave function away from the ground state.
Assume that this deformation amounts to a change in the ground state's magnetic order, which results in a change of the system's spin density, $\Delta s_{i}(\mathbf{r}) \equiv s_{i}(\mathbf{r})-s_{0}(\mathbf{r})$, where $s_{0}(\mathbf{r})$ stands for the ground state spin density, while $s_{i}(\mathbf{r})$ stands for the new/deformed state's spin density.
In such a case, one can show (see Appendix~\ref{app:eq_rho-s}) that, to first order in the deformation, the resulting change in the charge density, $\Delta \rho_{i} (\mathbf{r}) \equiv \rho_{i}(\mathbf{r}) - \rho_{0}(\mathbf{r})$ (with $\rho_{0}$ and $\rho_{i}$ standing for the ground state and deformed state charge density), is proportional to the change in the spin density, 
\begin{eqnarray}
  \Delta \rho_{i}(\mathbf{r}) &\approx& \frac{\lambda}{w} \, X_{i}(\mathbf{r}) \, \Delta s_{i}(\mathbf{r}) \,. \label{eq:Deltairho}
\end{eqnarray}
In the above expression $\lambda$ is the coupling constant connecting the orbital and the spin levels, while $w$ is the energy scale associated with the orbital degrees of freedom, and $X_{i}(\mathbf{r})$ is a numerical factor related to the type of spin deformation. 
Thus the ratio $\Delta \rho_{i} (\mathbf{r}) / \Delta s_{i}(\mathbf{r})$ gives direct access to the magnitude of $\lambda/w$.

A simple way of estimating the magnitude of the coupling $\lambda/w$ is to compute the \textit{average charge-spin susceptibility}, $\chi_{cs}$, defined as \cite{Narayan_arXiv:2017}
\begin{eqnarray}
  \chi_{cs} &\equiv& \frac{1}{N} \, \sum_{i=1}^{N} \, \chi_{i} \,, \label{eq:chgsp1}
\end{eqnarray}
where $N$ stands for the number of different magnetic orders considered for each material (see online data \cite{onlinedata}).
$\chi_{i}$ stands for the pairwise charge-spin susceptibility of each magnetic order with respect to the ground state.
This is defined as
\begin{eqnarray}
  \chi_{i} &\equiv& \frac{\Delta \rho_{i}}{\Delta s_{i}} \,, \label{eq:chgsp1b}
\end{eqnarray}
where $\Delta \rho_{i}$ ($\Delta s_{i}$) stands for the spatial fluctuations in charge (spin) density relative to the lowest-energy magnetic state.
The former are given by
\begin{subequations} \label{eq:chgsp2}
\begin{eqnarray}
  \Delta \rho_{i} &=& \int \ud \mathbf{r} \, \big\vert \rho_{i}(\mathbf{r}) - \rho_{0}(\mathbf{r}) \big\vert \,, \\
  \Delta s_{i} &=& \int \ud \mathbf{r} \, \big\vert s_{i}(\mathbf{r}) - s_{0}(\mathbf{r}) \big\vert \,.
\end{eqnarray}
\end{subequations}
where $\rho_{0}(\mathbf{r})$ and $s_{0}(\mathbf{r})$ are the charge and spin distributions of the lowest-energy magnetic state.

\subsection{Calculating the charge-spin susceptibility}

We calculate $\chi_{cs}$ as defined in Eqs.~\ref{eq:chgsp1}-\ref{eq:chgsp2}, by generating several low-energy magnetic textures for each material. 
As represented in Fig. \ref{fig:scheme}, in order to obtain a new magnetic order we optimize initial magnetic textures that differ from the ground state's one by a few flipped transition metal atoms' magnetic moments.
Then we compute the charge density and spin density differences between the ground and the new state obtaining $\Delta \rho_{i}$ and $\Delta s_{i}$ from Eqs.~\ref{eq:chgsp2}.
With several different magnetic orders we compute the average charge-spin susceptibility $\chi_{cs}$ from Eqs.~\ref{eq:chgsp1} and \ref{eq:chgsp1b}.

Since our objective is to screen a large set of materials against the charge-spin susceptibility, we decided to base our search protocol on a low-cost but sufficiently accurate computational method.
With that in mind we chose Kohn-Sham density functional theory (KS-DFT) \cite{Kohn_PR:1965}.
Most of the calculations presented in this work were performed using the KS-DFT approach \cite{Kohn_PR:1965}, as implemented in the QUANTUM ESPRESSO code \cite{Giannozzi_JPCM:2009}.
The exchange-correlation energy was approximated by the generalized gradient approximation (GGA) using the Perdew-Burke-Ernzerhof (PBE) functional \cite{Perdew_PRL:1996}.
To improve the description of the $d$ orbitals, we used the DFT+U scheme of Cococcioni and de Gironcoli \cite{Cococcioni_PRB:2005}.
Interactions between valence and core electrons were described by pseudopotentials in the accurate set of the Standard Solid-State Pseudopotentials library \cite{Lejaeghere_Science:2016,Prandini_NPJcomp:2018}.
The Kohn-Sham orbitals were expanded in a plane-wave basis with a cutoff energy $E_{c}$ (Ry), while a cutoff of $4 E_{c}$ was used for the charge density (see online data for $E_{c}$ of each material \cite{onlinedata}).
The $E_{c}$ of a given compound was chosen to be the largest $E_{c}$ among those of its constituent chemical elements.
The $E_{c}$ of an atomic species was estimated from checking for convergence of the single-atom's total energy against $E_{c}$.
Convergence was assumed when total energy changed less than $0.01$ Ry upon an increase of $E_{c}$ by $10$ Ry.
The Brillouin zone (BZ) was sampled using a $\Gamma$-centered 6x6x6 grid following the scheme proposed by Monkhorst-Pack \cite{Monkhorst_PRB:1976}.
Total energy convergence against the BZ grid density was tested by doing 7x7x7 grid calculations for unpolarized and ferromagnetic textures.
The crystal structure for each material was set up with the information available on the ICSD database \cite{Hellenbrandt_CR:2004} -- see online data \cite{onlinedata} for the CIF(s) used in the calculations of each material.
A supercell was used whenever the material unit cell had less than 4 transition metal atoms per unit cell.
This ensures that we can generate sufficient magnetic textures to properly estimate the charge-spin susceptibility.

For each material, we performed multiple DFT+U calculations (with $U=0,5,10$~eV) in order to assess the uncertainty in the charge-spin susceptibility estimate.
With the aim of converging different magnetic orders, we performed calculations in which the self-consistent cycle started from different magnetic states (see Fig. \ref{fig:scheme}), i.e. different orderings and magnitudes for the magnetic moments on the transition metal atoms.
On \cite{onlinedata} the reader can find data specifying all the DFT+U calculations that were performed, including material name, crystallographic identifier (CIF), Hubbard U, $E_{c}$ cut-off, supercell size, k-point mesh, starting magnetic state, final magnetic state, band gap estimate and total energy.

To check the accuracy of the results obtained from the DFT+U calculations, in sub-Section \ref{sec:charge-spin_methods} we compare charge-spin susceptibilities obtained from DFT+U with those obtained from the PBE and PBE0 hybrid functionals \cite{Perdew_JCP:1996} (as implemented in the CRYSTAL17 code \cite{Dovesi_WIRCMS:2018}), as well as those obtained from the highly accurate fixed-node diffusion Monte Carlo (DMC) \cite{Foulkes_RMP:2001} (as implemented on the quantum Monte Carlo package QWALK \cite{Wagner_JCP:2009}).

Fixed node diffusion Monte Carlo (DMC) is a fully first-principles stochastic framework to solve the Schr\"odinger equation which yields a variational upper bound to the ground state \cite{Foulkes_RMP:2001}.
We employed a Slater-Jastrow trial wavefunction, as implemented in the QWalk package \cite{Wagner_JCP:2009}.
We constructed the Slater determinant with orbitals from DFT calculations using the Crystal code \cite{Dovesi_WIRCMS:2018} employing the PBE0 functional \cite{Perdew_JCP:1996}.
Previous studies have shown that per comparison to other commonly used DFT functionals, PBE0 typically gives the best wave function nodes \cite{Kolorence_PRB:2010,Zheng_PRL:2015,Busemeyyer_PRB:2016}.
The Brillouin zone (BZ) was also sampled using a Gamma-centered 6x6x6 Monkhorst-Pack grid \cite{Monkhorst_PRB:1976}.
We used Dirac-Fock pseudopotentials and ECPs specially constructed for quantum Monte Carlo computations \cite{Burkatzki_JCP:2007,Burkatzki_JCP:2008}.
We controlled finite-size errors by using $2 \times 1 \times 1$ supercells and averaging over the sampled k-points.
We used a timestep of 0.005 Ha$^{-1}$.
This setup has been shown to give a good description of challenging materials like the cuprates \cite{Wagner_PRB:2015} and FeSe \cite{Busemeyyer_PRB:2016}.

For each material in sub-Section \ref{sec:charge-spin_methods} we considered several magnetic textures.
Some of these have zero total spin projection $S_{z}=0$ (i.e. antiferromagnetic-like orders) while others have $S_{z} \neq 0$ (ferromagnetic or flip orders).
The data sets provided online \cite{onlinedata} identify the magnetic textures considered for each material.
There we have also included figures representing those magnetic textures \cite{onlinedata}.

\section{Results and discussion}

\label{sec:results}
\begin{figure*}
  \centering
  \includegraphics[width=0.8\textwidth]{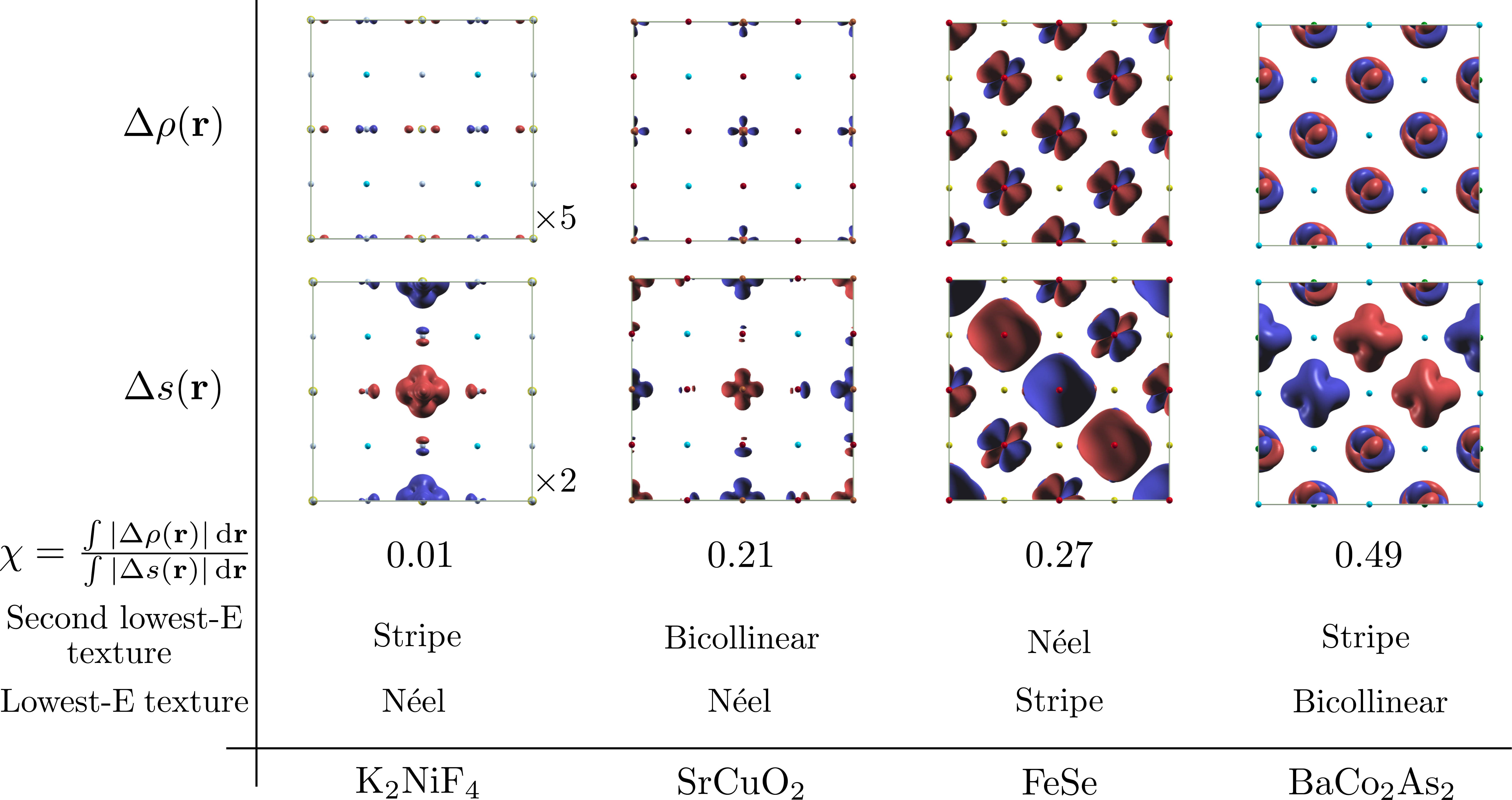}
  \caption{\textbf{DFT+U(5 eV) charge (top row) and spin (bottom row) density differences in the ab plane of the two lowest-energy magnetic textures for K$_{2}$NiF$_{4}$, SrCuO$_{2}$, FeSe and BaCo$_{2}$As$_{2}$.}
  Atoms are represented by spheres as follows: Ni (yellow), F (cyan) and K (gray) for K$_{2}$NiF$_{4}$; Cu (brown), O (red) and Sr (blue) for SrCuO$_{2}$; Fe (red) and Se (yellow) for FeSe; Co (green), Ba (gray) and As (cyan) for BaCo$_{2}$As$_{2}$.
  The red and blue isodensity surfaces stand for positive and negative values.
  The charge and spin density difference surfaces in the panels corresponding SrCuO$_{2}$, FeSe and BaCo$_{2}$As$_{2}$ were drawn with isolevel $0.01~e/\mathring{A}^{3}$, while K$_{2}$NiF$_{4}$ is at $0.002~e/\mathring{A}^{3}$ for charge density and $0.005~e/\mathring{A}^{3}$ for spin density. 
   }
  \label{fig:Charge_and_spin_densities}
\end{figure*}

\subsection{Visualizing the charge-spin response}

\label{sec:charge-spin_smallset}

In Fig. \ref{fig:Charge_and_spin_densities} we plot the change in charge and spin density between the lowest- and second lowest-energy magnetic texture for K$_{2}$NiF$_{4}$, SrCuO$_{2}$, FeSe and BaCo$_{2}$As$_{2}$.
Stoichiometric K$_{2}$NiF$_{4}$ is a typical Mott insulator showing N\'eel order \cite{Birgeneau_PRB:1970}.
SrCuO$_{2}$ is a N\'eel ordered magnetic insulator, while FeSe is a Hund's metal without long-range magnetic order at atmospheric pressure.
SrCuO$_{2}$ and FeSe are representative of the families of cuprate and iron-based high-temperature unconventional superconductors, well known to support a wide range of uncommon phases, ranging from strange metallic behavior to non-trivial magnetic states and high-temperature unconventional superconductivity \cite{Proust_ARCMP:2019,Wang_NatComm:2016,Medvedev_NatMat:2009}.
BaCo$_{2}$As$_{2}$ is a disordered magnetic metal \cite{Anand_PRB:2014}.
In the bottom row of Fig. \ref{fig:Charge_and_spin_densities} we show the pairwise charge-spin susceptibility $\chi_{i}$ (see Eqs.~\ref{eq:chgsp1b} and \ref{eq:chgsp2}) resulting from those magnetic textures.

In the leftmost column of Fig. \ref{fig:Charge_and_spin_densities} we see that the charge density of K$_{2}$NiF$_{4}$ is just slightly rearranged when the magnetic texture is modified.
This weak charge density response to changes in magnetic order persists for other magnetic textures of K$_{2}$NiF$_{4}$, which indicates that charge and spin degrees of freedom are weakly coupled in this material.

In the remaining columns of Fig. \ref{fig:Charge_and_spin_densities} we can see that the charge density response in SrCuO$_{2}$, FeSe and BaCo$_{2}$As$_{2}$ is much stronger than that in K$_{2}$NiF$_{4}$ whose iso-charge density surfaces were 5-fold magnified with respect to those of the other materials.
However, the charge-spin response in SrCuO$_{2}$, FeSe and BaCo$_{2}$As$_{2}$ is rather different.
Both the way in which charge rearranges and the magnitude of that rearrangement varies across these materials, as can be inferred from the shape and size of the isodensity-difference surfaces in Fig. \ref{fig:Charge_and_spin_densities}.
For instance, changing the magnetic order from N\'eel (ground state) to bicolinear order in SrCuO$_{2}$ mostly results in electrons moving from oxygen $2p_{x}$ ($2p_{y}$) orbitals into copper $3d_{x^{2}-y^{2}}$ ones.
In FeSe, changing from the stripe (ground state) to the N\'eel order seems to largely transfer electrons from iron's 3d$_{xz}$, 3d$_{yz}$ and 3d$_{xy}$ orbitals into its 3d$_{z^{2}-r^{2}}$.
Similarly, the change from bicollinear to collinear magnetic order in BaCo$_{2}$As$_{2}$ mostly redistributes electrons among the 3d orbitals of cobalt.

The spin density differences for these four materials (see second row of Fig. \ref{fig:Charge_and_spin_densities}) have similar magnitudes even if they are qualitatively different.
This thus suggests that the charge-spin response in K$_{2}$NiF$_{4}$ should be much weaker than that in SrCuO$_{2}$ and FeSe, which in turn seem to have a somewhat weaker response than BaCo$_{2}$As$_{2}$.
Those observations are corroborated by the values of the pairwise charge-spin susceptibility $\chi_{i}$ calculated for these magnetic textures and shown in the third row of Fig. \ref{fig:Charge_and_spin_densities}.

\subsection{Charge-spin susceptibility from different methods}

\label{sec:charge-spin_methods}

\begin{figure}
  \centering
  \includegraphics[width=0.99\columnwidth]{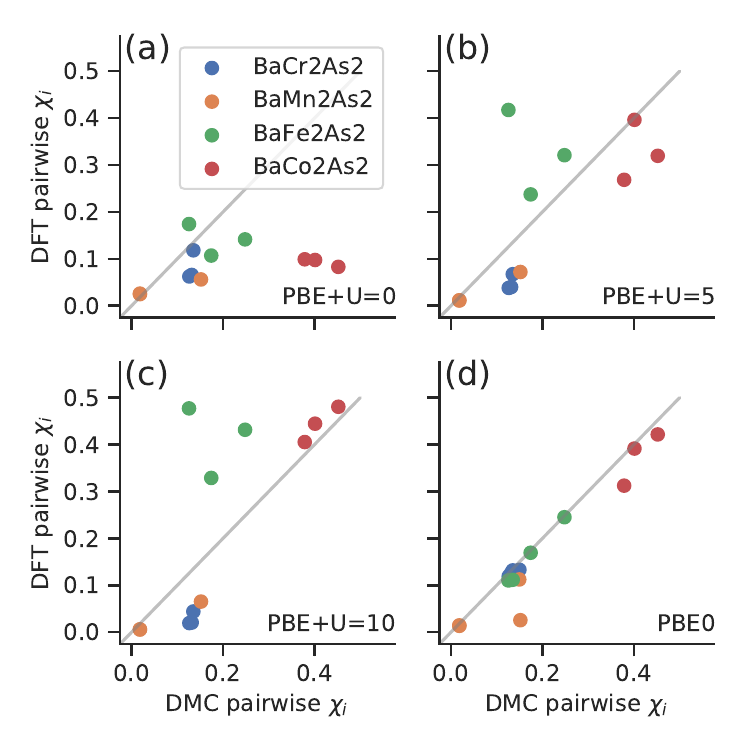}
    \begin{tabular}{c | c | c | c | c}
      \textrm{Functional} & \textrm{PBE+U=0} & \textrm{PBE+U=5} & \textrm{PBE+U=10} & \textrm{PBE0}  \\ [0.1ex]
      \hline
      \textrm{RMSD} & 0.17 & 0.11 & 0.14 & 0.04 \\ [0.1ex]
    \end{tabular}
  \caption{\textbf{Pairwise charge-spin susceptibility $\chi_{i}$ using different electronic structure methods.}
    Each point corresponds to the $\chi_{i}$ -- see Eqs.~\ref{eq:chgsp1} and \ref{eq:chgsp2} -- for an excited magnetic texture with respect to the ground state.
    The point colors identify the material in the set BaM$_{2}$As$_{2}$ with M = Cr, Mn, Fe, Co.
    The point's horizontal position is set by the diffusion Monte Carlo $\chi_{i}$ and the vertical one is set by the density functional theory $\chi_{i}$.
    Each panel corresponds to a different DFT functional: PBE+U=0,5,10 and PBE0.
    The table shows the root mean square deviation (RMSD) between the $\chi_{i}$ computed from a DFT method and those obtained with DMC.
  }
  \label{fig:ChgSp_methods}
\end{figure}

We now compare the charge-spin susceptibility computed from a few different methods: PBE+U \cite{Perdew_PRL:1996,Cococcioni_PRB:2005} with $U=0,5,10$~eV (with the plane-wave code \texttt{quantum espresso} \cite{Giannozzi_JPCM:2009}); PBE0 \cite{Perdew_PRL:1996} (using the localized basis code \texttt{CRYSTAL17} \cite{Dovesi_WIRCMS:2018}); and fixed-node diffusion Monte Carlo \cite{Foulkes_RMP:2001} (using the quantum Monte Carlo package \texttt{QWalk} \cite{Wagner_JCP:2009}).
Due to the high computational cost of the DMC calculations, we performed this comparison for a set of four barium arsenides with a ThCr$_{2}$Si$_{2}$-like structure: BaM$_{2}$As$_{2}$ with M = Cr, Mn, Fe, Mn.
In Fig. \ref{fig:ChgSp_methods} we compare the pairwise charge-spin susceptibility $\chi_{i}$ obtained with diffusion Monte Carlo (x-axis) with the $\chi_{i}$ obtained from density functional theory (y-axis).
Each panel in Fig. \ref{fig:ChgSp_methods} makes this comparison for DFT calculations done with each of the four functionals mentioned above: PBE+U=0,5,10 and PBE0.

In this figure the PBE+U derived pairwise charge-spin susceptibilities generally follow the trends of the more expensive PBE0 and DMC results.
The ordering of these four materials according to their values of pairwise charge-spin susceptibility $\chi_{i}$ resulting from DFT calculations with the PBE0 and DFT+U=5,10~eV functionals is the same as that resulting from DMC: $\chi_{Mn} \lesssim \chi_{Cr} \lesssim  \chi_{Fe} \lesssim \chi_{Co}$.

The deviation between the DFT-derived pairwise charge-spin susceptibilities and those calculated from DMC is shown in Fig. \ref{fig:ChgSp_methods}'s table.
PBE0 deviates the least from DMC, with a root mean square deviation of $\textrm{RMSD}_{\textrm{PBE0}} = 0.04$ considerably smaller than those resulting from the DFT+U calculations.
Part of the discrepancy between the PBE+U and PBE0/DMC $\chi_{i}$'s arises from the fact that some magnetic textures obtained with PBE+U are often quantitatively different from those obtained with PBE0/DMC.
Among the PBE+U functionals, U=5~eV is the one that better captures the response of charge to changes in the spin texture of these materials.
The PBE+U=5~eV $\chi_{i}$ root mean square deviation from DMC is $\textrm{RMSD}_{\textrm{U=5}} = 0.11$, while that of U=0~eV and U=10~eV is slightly larger: $\textrm{RMSD}_{\textrm{U=0}} = 0.17$ and $\textrm{RMSD}_{\textrm{U=10}} = 0.14$.

Even if the quantitative agreement between the DFT+U methods and DMC is not perfect, these still capture the qualitative trends in charge-spin susceptibility, enabling their use in charge-spin susceptibility screenings of large sets of materials.
In what follows we will show results for DFT+U with $U=5$~eV since this functional minimizes deviations from the DMC results -- see table of Fig. \ref{fig:ChgSp_methods}.

In some materials the charge-spin response is strongly dependent on the type of change in the magnetic texture.
That can be seen in Fig. \ref{fig:ChgSp_methods} where the pairwise charge-spin susceptibilities show a large spread for BaFe$_{2}$As$_{2}$ but a small one for BaCr$_{2}$As$_{2}$.
This is reminiscent of electron-phonon coupling physics, in which some phonons are more strongly coupled to the material's electronic degrees of freedom than others.
For simplicity we take the average (see Eq.~\ref{eq:chgsp1}) but have checked that different strategies do not affect the results in the next section.

\subsection{Charge-spin susceptibility}

\label{sec:charge-spin_results}

\begin{figure}
  \centering
  \includegraphics[]{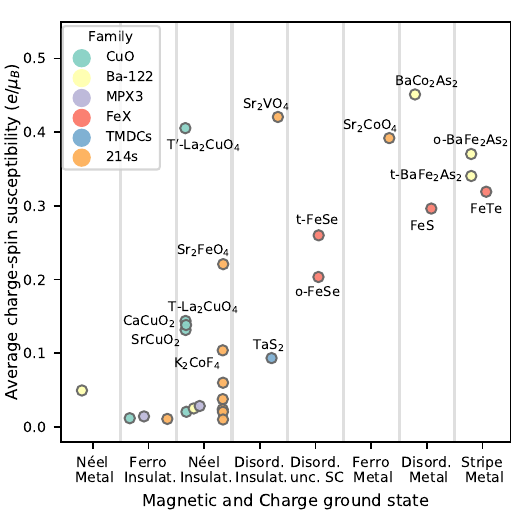}
  \caption{\textbf{Average charge-spin susceptibility for the test set materials.}
    Materials are positioned according to their low temperature magnetic and transport properties (x-axis) and their average charge-spin susceptibility (y axis) calculated with DFT+U with $U=5$~eV.
    Each point corresponds to one material.
    Those showing large charge-spin susceptibility are labeled, and tend to be materials with unconventional behavior.
  }
  \label{fig:ChargeSpin_ranking}
\end{figure}

\begin{figure}
  \centering
  \includegraphics[]{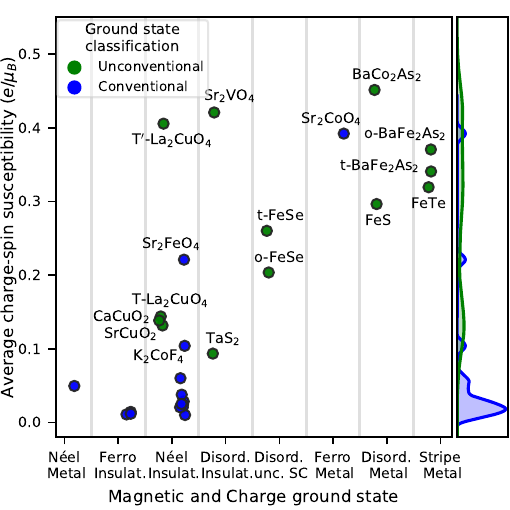}
  \caption{\textbf{Average charge-spin susceptibility for materials with conventional and unconventional ground states.}
    The x-axis position of materials is set according to their undoped and unpressured ground state.
    A material's average charge-spin susceptibility (from DFT+U with $U=5$~eV) determines the y-axis position.
    Each point corresponds to one material.
    The green (blue) colored points identify materials with unconventional (conventional) ground states.
    A material has unconventional ground states if its doping-pressure phase diagram shows unconventional superconductivity, disordered magnetic states or bad metal -- see Table \ref{tab:testset}.
    The right hand side panel shows the distribution of conventional (blue) and unconventional (green) materials in our test set according to their charge-spin susceptibility.
    Using a t-test, the populations are different with a p-value of $3\times10^{-5}$.
  }
  \label{fig:ChargeSpin_unconv-conv}
\end{figure}

In Fig.~\ref{fig:ChargeSpin_ranking} we show the charge-spin susceptibility for the materials in the entire test set from Table \ref{tab:testset} obtained using DFT+U with $U=5$~eV.
Each material is colored according to its family: copper oxides, barium arsenides, MPX$_{3}$'s, iron chalcogenides, transition metal dichalcogenides and 214 materials.
Overall, we find that materials showing large charge-spin susceptibility in Fig.~\ref{fig:ChargeSpin_ranking}, typically present unconventional properties either in their stoichiometric form, or when put under pressure or chemically doped.
This is shown in Fig.~\ref{fig:ChargeSpin_unconv-conv} where we color materials according to whether their pressure vs. doping phase diagram shows unconventional ground states.
Both these families are well separated: $\chi_{\textrm{conv}} = 0.08 \pm 0.10$ for conventional materials, while $\chi_{\textrm{unc}} = 0.28 \pm 0.12$ for unconventional ones.
Using a two-sided t-test, the populations have different means with a p-value of $3\times 10^{-5}$.
In this section, we discuss in some detail the materials in our test set showing large charge-spin susceptibility.

The results in Fig. \ref{fig:ChargeSpin_ranking} show four copper oxides with sizable values of charge-spin susceptibility: \textbf{T-La$_{2}$CuO$_{4}$}, \textbf{CaCuO$_{2}$} and \textbf{SrCuO$_{2}$} all have $\chi_{cs} \approx 0.15$ while \textbf{T'-La$_{2}$CuO$_{4}$} shows $\chi_{cs} \approx 0.40$.
All these four materials are N\'eel ordered \cite{Proust_ARCMP:2019,Lombardi_PRB:1996,Zaliznyak_PRL:2004,Jin_Nature:2011} insulators and well known to become unconventional superconductors under chemical doping \cite{Proust_ARCMP:2019,Lombardi_PRB:1996,Smith_Nat:1991,Jin_Nature:2011}.
Several other uncommon phases have been observed in these materials, ranging from a pseudogap phase, to strange metallicity, short-range magnetic and charge order \cite{Proust_ARCMP:2019,Lombardi_PRB:1996,Zaliznyak_PRL:2004,Jin_Nature:2011}.

Two other members of the copper oxide family, \textbf{TeCuO$_{3}$ and SeCuO$_{3}$}, show very small charge-spin susceptibility.
TeCuO$_{3}$ is a N\'eel ordered insulator while SeCuO$_{3}$ is a ferromagnetic insulator \cite{Lawes_PSSC:2009}, both well known to show magnetodielectric properties \cite{Lawes_PSSC:2009}.
As opposed to the four cuprates with copper-oxide planes discussed above, unconventional phases have not been observed in either TeCuO$_{3}$ or SeCuO$_3$, consistent with the small value of susceptibility computed here.

The iron pnictides and chalcogenides \textbf{o-FeSe}, \textbf{t-FeSe}, \textbf{FeS}, \textbf{FeTe}, \textbf{t-BaFe$_{2}$As$_{2}$} and \textbf{o-BaFe$_{2}$As$_{2}$}, all show $\chi_{cs} \approx 0.30-0.37$.
These materials are all metallic magnets, some showing stripe magnetic textures (BaFe$_{2}$As$_{2}$ and FeTe \cite{Wang_PRL:2009,Martinelli_PRB:2010}) while others only present short-range order (FeSe and FeS \cite{Holenstein_PRB:2016,Wang_NatComm:2016}).
All these materials have unconventional superconducting phases induced by pressure or doping \cite{Rotter_PRL:2008,Sales_PRB:2009,Lai_JACS:2015,Medvedev_NatMat:2009}, as well as short-range magnetic order and bad metal phases \cite{Luo_PR:2012,Kasahara_PRB:2010,Martinelli_PRB:2010,Sales_PRB:2009,Wang_NatComm:2016}.

\textbf{Sr$_{2}$FeO$_{4}$} has $\chi_{cs}=0.22$ and is an antiferromagnetic semiconductor.
Chemically doping this compound weakens both its antiferromagnetic ordering and semiconducting character without completely suppressing the electronic gap \cite{Omata_PRB:1994,Jennings_MCP:2005,Zhao_JCC:2011}.
It has been shown that pressure induces a semiconductor-to-metal transition at $P\approx18$ GPa \cite{Rozenber_PRB:1998}, but so far no unconventional phases have been observed on its phase diagram down to $5$ K for pressures up to $30$ GPa.

\textbf{Sr$_{2}$CoO$_{4}$} is ferromagnetic and metallic at low temperatures \cite{Wang_PRB:2005}.
Upon chemical doping with Y \cite{Wang_PRB:2005}, La \cite{Shimada_PRB:2006} and Nd \cite{Yao_JAP:2012} the ferromagnetism weakens and semiconducting behavior arises.
This material becomes a superconductor at around 5K when doped with H$_2$O \cite{takada_superconductivity_2003}, with very similar properties to the cuprate superconductors. 
We were not aware of this result prior to the study; the material was identified purely due to the charge-spin descriptor, $\chi_{cs}=0.39$.

 \textbf{Sr$_{2}$VO$_{4}$} is a multi-orbital Mott insulator with no long-range magnetic order\cite{Yamauchi_PRB:2015} that can be driven into a metallic state by hydrostatic pressure ($\approx 20 - 24$ GPa) \cite{Karmakar_PRL:2015}.  
An unconventional metal emerges at low temperatures in the vicinity of the pressure-driven transition \cite{Karmakar_PRL:2015}.
In contrast to Sr$_{2}$VO$_{4}$ thin films \cite{Matsuno_APL:2003}, attempts to chemically dope the bulk crystal did not succeed in making it metallic \cite{Deslandes_PhysC:1991}.
Since our calculations suggest it has a strong charge-spin coupling, $\chi_{cs}=0.42$, a more comprehensive exploration of different ways of chemically doping this material may reveal novel phases.

\textbf{BaCo$_{2}$As$_{2}$} has a rather large charge-spin susceptibility, $\chi_{cs}=0.45$.
It is a disordered magnetic metal \cite{Anand_PRB:2014} that seems to remain so upon both chemical doping with K \cite{Anand_PRB:2014} and hydrostatic pressure (up to $8$ GPa) \cite{Ganguli_MRB:2013}.
It thus seems similar to Sr$_2$VO$_4$ in the sense that there is a disordered magnetic state.
We note that these two systems are not ordinary magnetic materials, and both exhibit nonstandard ground states. 
Thus the charge-spin descriptor succeeded in identifying unusual physics in these materials.

\textbf{K$_{2}$CoF$_{4}$} has been classified as a 2D Ising magnet \cite{Breed_Physica:1969} owing to its strongly anisotropic magnetic interactions.
To our knowledge, this material's behavior under pressure or chemical doping has been very sparsely studied \cite{Breed_JAP:1970}, potentially due to the presence of fluorine.
Its charge-spin susceptibility, $\chi_{cs}=0.10$, is slightly below that of some superconducting cuprates.

\textbf{TaS$_{2}$} is a Mott insulator associated with a charge-density wave (CDW) phase \cite{Sipos_NatMat:2008}.
Pressure induces a metal-to-insulator transition and a superconducting state below $5$ K \cite{Sipos_NatMat:2008}.
Recent experiments \cite{Kratochvilova_NPJQM:2017} suggest that the low-temperature CDW phase has short-range magnetic order, which supported proposals that this material might realize a quantum spin liquid state \cite{Law_PNAS:2017}.
Our calculations, done with the undistorted crystal structure, give $\chi_{cs}=0.09$ just below what we get for some cuprates.

The compounds with lower charge-spin susceptibility in Fig.~\ref{fig:ChargeSpin_unconv-conv} (listed on the bottom of Table \ref{tab:testset}) comprise materials that show conventional phases, mostly insulators with N\'eel antiferromagnetic order and ferromagnetic order.
Some of these materials are known to become metallic (eg. BaMn$_{2}$As$_{2}$) or to acquire spin and/or charge stripe order (eg. La$_{2}$NiO$_{4}$ and La$_{2}$CoO$_{4}$) upon doping or pressure, but none of them presents unconventional electronic phases.
This indicates that, as suggested by our calculations, their charge and spin degrees of freedom are weakly coupled.

\section{Conclusion}

\label{sec:conclusion}

We have presented a new way of probing charge-spin coupling in materials.
It is based on the charge-spin susceptibility, a quantity that estimates the magnitude of the coupling between charge and spin degrees of freedom in a material.
This quantity is straightforward to compute in a high-throughput workflow and when applied to a collection of layered materials containing transition metal atoms suggests that materials with high charge-spin coupling exhibit unconventional phenomena.

All the materials in our test set known to present unconventional phases (green colored in Fig.~\ref{fig:ChargeSpin_unconv-conv}) show average charge-spin susceptibilities $\chi_{cs} \geq 0.09$.
Among the 16 materials with charge-spin susceptibility above this value only three have not been observed to show unconventional physics, \ce{Sr2CoO4}, \ce{Sr2FeO4} and \ce{K2CoF4}, which is a rather small rate of false positive identifications.
Perhaps more importantly, the false negative rate was zero within our test set.
This rate is certainly good enough to motivate experimental investigation into materials. 

The computation of the charge-spin susceptibility for the materials in our test set was performed using DFT+U, a low-cost method which is sufficiently accurate to highlight the same qualitative trends found using the highly accurate many-body method fixed-node diffusion Monte Carlo.
These calculations are inexpensive enough that they can be used in large-scale probes of the strength of the coupling between electronic and magnetic degrees of freedom in materials.
The trends found here suggest that the charge-spin susceptibility is a valuable quantity for computational searches for new unconventional ground states, including unconventional superconductivity similar to iron-based and cuprate superconductors.

\begin{acknowledgments}
  This work was supported by the Center for Emergent Superconductivity, an Energy Frontier Research Center funded by the U.S. Department of Energy, Office of Science, Office of Basic Energy Sciences under Award Number DEAC0298CH1088.
  L.K.W. was supported by a grant from the Simons Foundation as part of the Simons Collaboration on the many-electron problem.
  The authors thank Daniel Shoemaker for many illuminating discussions.
  The computational resources used in this work were provided by the University of Illinois Campus Cluster and the Blue Waters sustained-petascale computing project, which is supported by the National Science Foundation (awards OCI-0725070 and ACI-1238993) and the state of Illinois. Blue Waters is a joint effort of the University of Illinois at Urbana-Champaign and its National Center for Superconducting Applications.
  This work's data is available through the Materials Data Facility \cite{Blaiszik_JOM:2016,Blaiszik_MRS:2019}.
\end{acknowledgments}




\appendix

\section{Charge density response and charge-spin coupling}

\label{app:eq_rho-s}

In the context of a system governed by the Hamiltonian in Eq.~\ref{eq:H}, assume that we fix the portion of the wave function associated with the spin degrees of freedom, $\phi(\mathbf{r})$, to a particular magnetic order.
Then, the orbital degrees of freedom will be described by $\varphi(\mathbf{r};\phi)$.
The charge and the spin density of such configuration will be given by
\begin{subequations}
\begin{eqnarray}
  \rho(\mathbf{r}) &=& \sum_{\sigma} \Big( \vert \varphi_{\sigma}(\mathbf{r}) \vert^{2} + \vert \phi_{\sigma}(\mathbf{r}) \vert^{2} \Big) \,, \\
  s(\mathbf{r}) &=& \sum_{\sigma}\sigma \Big( \vert \varphi_{\sigma}(\mathbf{r}) \vert^{2} + \vert \phi_{\sigma}(\mathbf{r}) \vert^{2} \Big) \,,
\end{eqnarray}
\end{subequations}
where $\sigma = \pm$ identifies the spin projection, while $\varphi_{\sigma}$ is a short-hand for $\varphi_{\sigma}(\mathbf{r};\phi)$.

Let us apply an infinitesimal deformation away from the ground state on the portion of the wave function describing the spin degrees of freedom, $\delta \phi_{\sigma}(\mathbf{r}) = \phi_{\sigma}(\mathbf{r}) - \phi_{0 \sigma}(\mathbf{r})$ (where $\phi_{\sigma}$ is the deformed wave function and $\phi_{0 \sigma}$ is the ground state one).
We can then write the differential of each component of the charge and spin densities as follows:
\begin{subequations} \label{eq:linear}
\begin{eqnarray}
  \delta \vert \phi_{\sigma}\vert^{2} &=& 2 \, \vert \phi_{\sigma} \vert \, \delta \vert \phi_{\sigma} \vert \,, \\
  \delta \vert \varphi_{\sigma}\vert^{2} &=& 2 \, \vert \varphi_{\sigma} \vert \, \sum_{\alpha} \frac{\delta \vert \varphi_{\sigma} \vert}{\delta \vert \phi_{\alpha} \vert} \, \delta \vert \varphi_{\alpha} \vert \,,
\end{eqnarray}
\end{subequations}
where we used $\delta \vert \varphi_{\sigma} \vert = \sum_{\alpha} \frac{\delta \vert \varphi_{\sigma} \vert}{\delta \vert \phi_{\alpha} \vert} \delta \vert \phi_{\alpha} \vert$ under the assumption that the wave function's orbital degrees of freedom component only depend on the absolute value of the spin states component, $\varphi_{\sigma}(\mathbf{r}; \phi_{\sigma}) \simeq \varphi_{\sigma}(\mathbf{r}; \vert \phi_{\sigma} \vert)$.

We can write the differential of the charge and spin densities as
\begin{subequations}
\begin{eqnarray}
  \delta \rho &=& 2 \, \sum_{\sigma, \alpha} \bigg( \vert \varphi_{\sigma} \vert \, \frac{\delta \vert \varphi_{\sigma} \vert}{\delta \vert \phi_{\alpha} \vert} + \vert \phi_{\sigma}\vert \, \delta_{\sigma \alpha} \bigg) \, \delta \vert \phi_{\alpha} \vert \,, \\
  \delta s &=& 2 \, \sum_{\sigma, \alpha} \sigma \bigg( \vert \varphi_{\sigma} \vert \, \frac{\delta \vert \varphi_{\sigma} \vert}{\delta \vert \phi_{\alpha} \vert} + \vert \phi_{\sigma}\vert \, \delta_{\sigma \alpha} \bigg) \, \delta \vert \phi_{\alpha} \vert \,,
\end{eqnarray}
\end{subequations}
where $\delta_{\sigma \alpha}$ is the Kronecker delta.

Consider now that the small deformation is such that it only changes the spin states' magnetic order, preserving their contribution to the charge density, i.e. $\delta \sum_{\sigma} \vert \phi_{\sigma} \vert^{2} \simeq 0$.
This implies that $\sum_{\sigma} 2 \, \vert \phi_{\sigma} \vert \, \delta \vert \phi_{\sigma} \vert \simeq 0$, which allows us to write
\begin{eqnarray}
  \delta \vert \phi_{-} \vert &=& - \frac{\vert \phi_{+} \vert}{\vert \phi_{-} \vert} \delta \vert \phi_{+} \vert \,. \label{eq:delta_phi_pm}
\end{eqnarray}

Under this approximation we can write $\delta \rho$ and $\delta s$ as
\begin{subequations}
\begin{eqnarray}
  \delta \rho &=& \frac{2 \lambda}{w} \, \bigg( \Upsilon_{+} - \frac{\vert \phi_{+} \vert}{\vert \phi_{-} \vert} \Upsilon_{-} \bigg) \, \delta \vert \phi_{+} \vert \,, \\
  \delta s &=& 2 \, \bigg( \Xi_{+} + \frac{\vert \phi_{+} \vert}{\vert \phi_{-} \vert} \Xi_{-} \bigg) \, \delta \vert \phi_{+} \vert \,,
\end{eqnarray}
\end{subequations}
where $\Upsilon_{\alpha} \equiv \sum_{\sigma} \vert \varphi_{\sigma} \vert \, f_{\sigma \alpha}$ and $\Xi_{\alpha} \equiv \frac{\lambda}{w} \, \Upsilon_{\alpha} + \vert \phi_{\alpha} \vert$.
In these expressions we used the definition $\frac{\delta \vert \varphi_{\sigma} \vert}{\delta \vert \phi_{\alpha} \vert} \equiv \frac{\lambda}{w} \, f_{\sigma \alpha}$, where according to the main text's notation, $\lambda$ stands for the coupling between the spin states and the orbital degrees of freedom, while $w$ corresponds to a material-specific energy scale.

Using the above expressions we can write $\delta \rho$ in terms of $\delta s$ as
\begin{eqnarray}
  \delta \rho(\mathbf{r}) &=& \frac{\lambda}{w} \, \frac{\Upsilon_{+} - \frac{\vert \phi_{+}\vert}{\vert \phi_{-}\vert} \Upsilon_{-}}{ \Xi_{+} + \frac{\vert \phi_{+}\vert}{\vert \phi_{-}\vert} \Xi_{-} } \, \delta s(\mathbf{r}) \,. \label{eq:rho-s}
\end{eqnarray}
which using Eq.~\ref{eq:delta_phi_pm} can be simplified into
\begin{eqnarray}
  \delta \rho(\mathbf{r}) &=& \frac{\lambda}{w} \, X(\mathbf{r}) \, \delta s(\mathbf{r}) \,. \label{eq:rho-s-2}
\end{eqnarray}
where $X(\mathbf{r}) = \Upsilon_{+}(\mathbf{r}) / \vert \phi_{+}(\mathbf{r}) \vert$.

Eq.~\ref{eq:rho-s-2} has the same form of main text's Eq.~\ref{eq:Deltairho}, only that in the latter we use a perturbation theory notation: $\Delta s_{i}(\mathbf{r}) = s_{i}(\mathbf{r}) - s_{0}(\mathbf{r})$ [instead of $\delta s(\mathbf{r})$] and $\Delta \rho_{i}(\mathbf{r}) = \rho_{i}(\mathbf{r}) - \rho_{0}(\mathbf{r})$ [instead of $\delta \rho(\mathbf{r})$], where $\rho_{i}$ and $s_{i}$ stand for the charge and spin densities of the $i$ deformation away from the ground state's charge and spin densities, $\rho_{0}(\mathbf{r})$ and $s_{0}(\mathbf{r})$.

\bibliographystyle{apsrev4-1}

\end{document}